\documentclass[aps,tightenlines,twocolumn,superscriptaddress]{revtex4-1}
\usepackage{graphicx}% Include figure files
\usepackage{dcolumn}% Align table columns on decimal point
\usepackage{bm}% bold math
\usepackage{csquotes}
\usepackage{epigraph}
\usepackage{amssymb}
\usepackage{color,graphicx}
\usepackage{subfigure}
\usepackage{lmodern}
\usepackage{amsmath}
\usepackage{amsbsy}
\usepackage{amsthm}
\usepackage{float}
\usepackage{bbm}
\begin{document}

\title{Mimicking Classical Noise in Ion Channels by Quantum Decoherence}
\author{Mina Seifi}
\affiliation{Research Group on Foundations of Quantum Theory and Information,
Department of Chemistry, Sharif University of Technology
P.O.Box 11365-9516, Tehran, Iran}
\author{Ali Soltanmanesh}
\affiliation{Research Group on Foundations of Quantum Theory and Information,
Department of Chemistry, Sharif University of Technology
P.O.Box 11365-9516, Tehran, Iran}
\affiliation{Philosophy of Science Group, Sharif University of Technology, Tehran, Iran}
\author{Afshin Shafiee*}
\affiliation{Research Group on Foundations of Quantum Theory and Information,
Department of Chemistry, Sharif University of Technology
P.O.Box 11365-9516, Tehran, Iran}
\begin{abstract}
 The mechanism of selectivity in ion channels is still an open question in biology. According to recent proposals, it seems that the selectivity filter of the ion channel, which plays a key role in the channel's function, may show quantum coherence, which can play a role in explaining the selection mechanism and conduction of ions. However, due to decoherence theory, the presence of environmental noise causes decoherence and loss of quantum effects. Sometimes we hope that the effect of calssical noise of the environment in ion channels can be modeled through a picture whose the quantum decoherence theory presents. In this paper,  we simulated the behavior of the ion channel system in the Spin-Boson model using the unitary evolution of a stochastic Hamiltonian operator under the classical noise model. Also, in a different approach, we modeled the system evolution as a two-level Spin-Boson model with tunneling interacting with a bath of harmonic oscillators, using decoherence theory. The results of this system were discussed in different classical and quantum regimes. By examining the results it was found that the Spin-Boson model at a high hopping rate of Potassium ions can simulate the behavior of the system in the classical noise approach. This result is another proof for the fact that ion channels need high speed for high selectivity.  \end{abstract}

\maketitle
%%%%%%%%%%%%%%%%%%%%%%%%%%%%%%%%

\section{Introduction}

At first, it was believed that quantum phenomena such as entanglement or quantum tunneling do not exist in biological environments. Because these environments are hot, humid and noisy, it is expected that fast decoherence will occur in them and quantum features will be suppressed \cite{Tus,Qas,Moh}. New evidence has determined that quantum principles are necessary to explain some biological phenomena such as quantum effects in the brain, electromagnetic and rotational navigation of migratory birds, charge transfer through DNA, and photosynthesis \cite{Mar,Gha,Kim,Tir,Sch,Lam,Eng,Mer,Gie,Slo}. Accordingly, the role of quantum phenomena in the vital activities of living cells is accepted \cite{Ghas}. 

\begin{figure}
\centering
\includegraphics[scale=0.37]{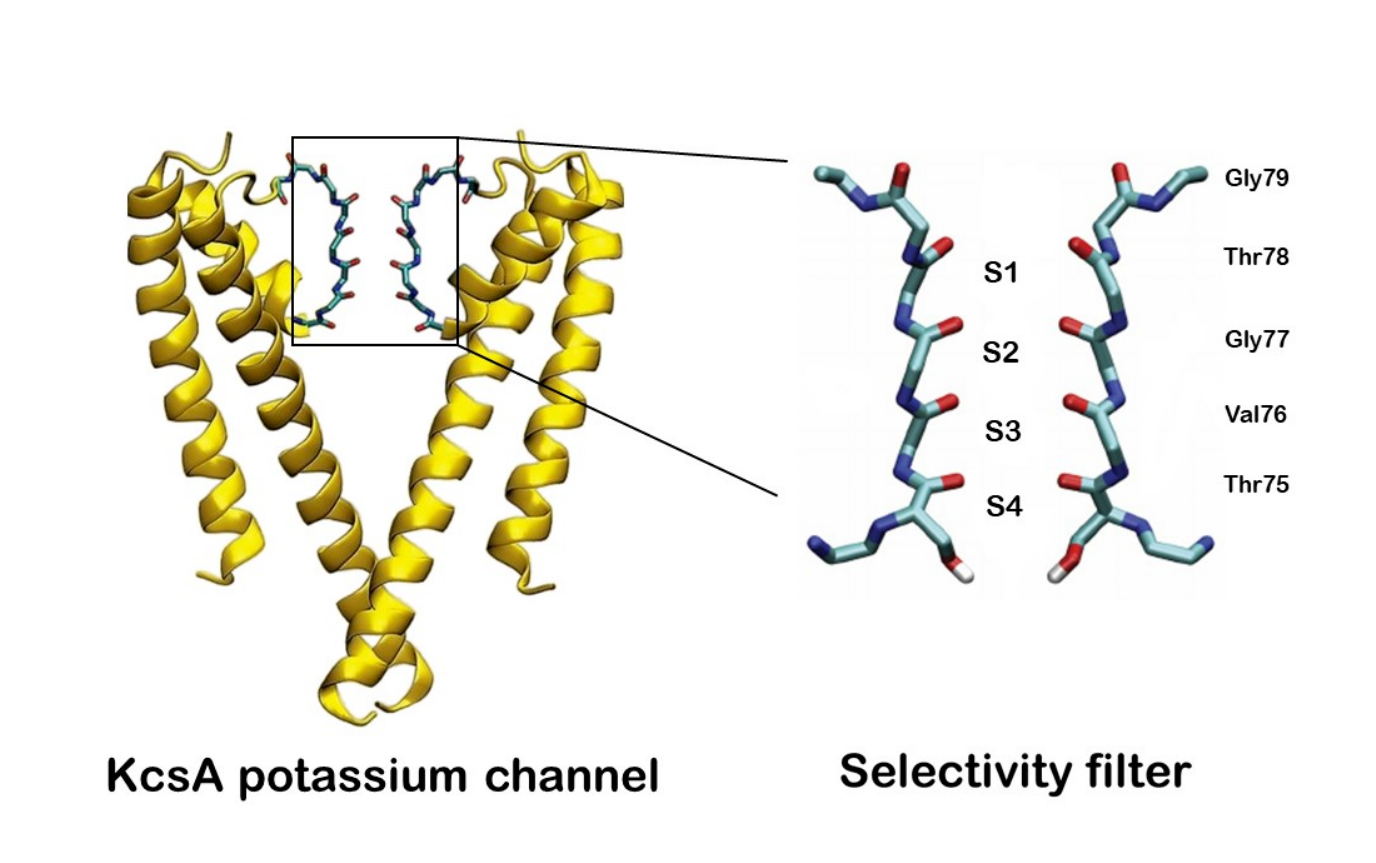}
\caption{(Left) A KcsA potassium ion channel representation after PDB 1K4C. (Right) Two P-loop monomers in the selectivity filter, composed of the sequences of TVGYG amino acids [T (Threonine, Thr75), V (Valine, Val76), G (Glycine, Gly77), Y (Tyrosine, Tyr78), G (Glycine, Gly79)].
}
\label{ion}
\end{figure}

Ion channels are a set of proteins embedded in the cell membrane, in which quantum effects may play a functional role based on the time, size, and energy scale of the transfer phenomena \cite{Gani}. Recently, it has been suggested that quantum coherence probably plays a role in the selectivity and transport of ions in these channels \cite{Sei,Vaz,Sal,Sum,Summ}. These channels regulate the flux of certain ions across the membrane \cite{Cor}. Different ion channels play an essential role in many biological processes, including muscle contraction, nerve signaling, cell proliferation, epithelial fluid transportation, and cell homeostasis \cite{Par,West,Gra,Shu}. Structurally, these membrane protein complexes are composed of several subunits whose spatial arrangement is such that they form sub-nanometer holes for the entry and exit of ions from the cell \cite{Tus}. One of the common features among ion channels is the presence of a selectivity filter whose task is to pass only a specific type of ion \cite{Nos,Sok,Ber}. The structure of this filter has been studied in the bacterial channel of Streptomyces lividans (KcsA) \cite{Doyl}. The selectivity filter (SF) is composed of four P-loop strands and has a structure similar to that of alpha-helix. Each P-loop is composed of five amino acids-Theronine (Thr75), Valine (Val76), Glycine (Gly77), Tyrosine (Tyr78), and Glycine (Gly79)- linked by peptide units (H–N–C=O), where N–C=O is an amide group and C=O is a carbonyl group. Carbonyls are responsible for trapping and displacing ions within the filter. The structure of the KcsA channel and selectivity filter is shown in FIG. \ref{ion}. Although the difference in the atomic radius of $\text{K}^{+}$ and $\text{Na}^{+}$ is only about 0.38 nm, potassium ions penetrate at least 10,000 times more than sodium ions through the selectivity filter \cite{All,Vah}. Potassium channels conduct $\text{K}^{+}$ ions with a speed close to the diffusion limit ($10^{6}$-$10^{8}$ $\text{s}^{-1}$) across the cell membrane \cite{Sala,Tri,Mora}.

Despite the extensive theoretical and experimental research, most biomolecular methods cannot well explain the ion selection at the nanoscale, and still, numerous issues remain unsolved \cite{Nta,Dud,Tho,Alle,Big,Dom,Shr,Rou,Bury,Chun,Ben}.
 An important paradox of ion channels is how a flexible structure such as a selectivity filter can select ions at high speed. This rapid and high selection is vital for the physiology of living organisms. After determining the atomic resolution structure of the bacterial KcsA channel, it seemed that classical models such as the snug-fit model could not accurately explain the ion selectivity process \cite{Nos,MacK,Nosk,Jia}. Since classical mechanics is not enough to explain this process in detail, and experimental evidence has determined the existence of quantum effects in biological cells, it seems that quantum mechanics should be used to solve this problem. To understand ion selectivity and transport in ion channels, accurate atomic models are needed that can well describe microscopic interactions. In other words, coherence probably plays a role in ion selection \cite{Vaz}. However, because the interaction between the system and its environment is strong, this coherence is probably short-term. One can only speculate about the functional roles of coherence in ion channels, in which, there is still no precise experimental technique to provide information about the parameters used in effective models. The issue of the regime (quantum or classical) where these channels can be modeled best, is a question that must be answered with the help of experience. 
 There are different theoretical frameworks for examining coherence in quantum systems \cite{Cif}. A standard approach is to model the system and the environment as a large quantum system and track the effects of the environment (usually in the Born-Markov approximation) to obtain a single description of the system in question. An alternative approach is that instead of considering the system and its environment completely quantum, they consider the effect of the environment as a classical noise in the degrees of freedom of the system \cite{Bing}. In this picture, quantum decoherence is simulated by averaging over a set of random but unitary quantum dynamics. This approach can often mimic standard open system dynamics and be used to simulate environmental effects. It should be noted that there is a fundamental difference between decoherence and classical noise, and this difference is well-understood. In real decoherence, the unitary evolution of the system and the environment causes the non-unitary evolution of the reduced density matrix of the system. In the classic noise model, because the system is not coupled with any external environment, the dynamics of the system in each individual realization of the noise is unitary, so the loss of coherence is determined only in an ensemble. Despite the fundamental difference between classical noise and decoherence, the noise model can mimic the effects of decoherence well because both models reduce coherence. The stochastic picture with classical noise has been widely used in physics and chemistry for various cases such as non-Markovian dynamics, many-body open quantum systems, and central spin problems \cite{Cost,Aur,Yan}. Budini investigated the effect of classical random fields on quantum systems, specifically for studying the heating of trapped ions. He used a stochastic Hamiltonian for the system and by using averaged dynamics, he obtained a general description of the decoherence behavior of the system \cite{Bud}. Chenu \textit{et al.} presented a scheme for many-body  decoherence simulation based on the unitary evolution of a stochastic Hamiltonian due to the addition of classical noise \cite{Aur}. Classical noise facilitates the experimental realization of such a simulation. Szańkowski and Cywiński formulated the necessary criteria for simulating the dynamics of open quantum systems with an external noise field that replaces the environmental degrees of freedom \cite{Noi}. Crow and Joynt showed that the decoherence caused by the interaction of a qubit and a quantum bath can be classically simulated by the unitary evolution of a stochastic Hamiltonian \cite{Cro}. Gu and Franco compared the quantum decoherence, which is caused by the entanglement of the system with its environment, with the apparent decoherence, which is obtained by averaging over an ensemble of unitary evolutions caused by a stochastic Hamiltonian, and presented the necessary conditions for quantitative modeling of decoherence by the classic noise picture \cite{Bing}. Ma \textit{et al.} used a Gaussian noise model for electron spin qubits in natural silicon and showed a good agreement between the results of the classical noise model with fully-quantum bath theory and experimental measurements to investigate the decoherence of this system \cite{Ma}. The dynamics of quantum systems under quantum and classical noise were investigated by Saira \textit{et al.} Quantum noise arises from the coupling of the microscopic system to its macroscopic environment, while classical noise is described by a random process that causes the time evolution of a closed quantum system. According to this analysis, fully quantum models can be depicted in the Born approximation with a quantum system under classical noise \cite{Sa}. Schneider and Milburn considered white noise as random processes of the decoherence source for a trapped ion and presented a simple master equation of this source. The results of their investigation showed a good agreement with the recent experiments in terms of quality \cite{Ge}.
 
 In this work, the focus is on whether solving the ion channel problem considering the thermal bath is a correct assumption or not. The main question is ’When can classical noise be mimicked with quantum decoherence for potassium ion channel?’ This work has been done by examining the system in the presence of environmental noise and comparing the results with real decoherence. To investigate the quantum decoherence in this system, the Spin-Boson model has been used. Classical noise is modeled with a stochastic parameter in the Hamiltonian.  Then, conditions are analyzed under which the quantum coherence dynamics of an ion channel can be simulated by a stochastic Hamiltonian operator, which is an alternative to environmental degrees of freedom.
 
 The present paper is organized as follows. In Section II, we introduce the formalism of environmental noise and explain the resulting loss of coherence. In Section III, we review the decoherence theory and specifically the Spin-Boson model. In Section IV, we introduce the desired model for the ion channel and with the help of two Spin-Boson and classical noise models, we check the loss of coherence in our purpose system. Then, the results of the previous two sections are compared in detail in Section V. Finally, the paper in Section VI is concluded.
%%%%%%%%%%%%%%%%%%%%%%%%%

\section{Classical noise formalism }
There are different strategies for modeling the dynamics of open quantum systems \cite{Bre}. In this part, the classical noise approach is introduced for mimicking decoherence, whereby the Hamiltonian includes a dynamical stochastic variable. The effect of the environment in this approach is to introduce classical noise in the degrees of freedom of the system \cite{Bing,Cost,Aur,Yan,Mon}. The fundamental difference between classical noise and decoherence is that in the decoherence mechanism, the system is entangled with the environment, and the phase relations are irreversibly lost from the system. By contrast, in classical noise representation, the entanglement of the environment system does not occur because the entanglement is a type of non-classical correlation \cite{Joos}. In classical noise formalism, the decay of coherence can only be manifest in an ensemble. Because the system is not coupled with any environment, the dynamics of the system are completely unitary in each individual realization of the noise process. So the decay of coherence appears by averaging over an ensemble of stochastic but unitary dynamics.  Since both noise and decoherence models effectively cause the loss of coherence, the decoherence model can mimic the classical noise \cite{Bing}. However, the fundamental conceptual differences between the two still remain. Classical noises are generally classified into Gaussian and non-Gaussian according to the form of their probability distribution. Gaussian noises are the most common and simplest types of noises. Usually, any classical noise is specified by its autocorrelation or equivalent power distribution (noise spectrum) \cite{Yan}. For Gaussian environments (bosonic and fermionic environments), the noise is completely determined by the autocorrelation function. Non-Gaussian noises are usually used for anharmonic environments (spin environments), and multi-time correlation functions are needed to characterize these noises. Recently, Kiely has modeled noise in quantum systems using a dynamic stochastic parameter in the Hamiltonian system \cite{Kie}. He investigated the derivation of exact master equations for several different common noises. Using these exact master equations avoids the numerous iterations needed to obtain accurate mean dynamics.  Specifically, the exact master equation for colored Gaussian noise is as follows \cite{Cost,Budi}:

\begin{align}
\label{mastereq}
\frac{d}{dt}\hat{\rho}=-\dfrac{i}{\hbar}[\hat{H}_{0},\hat{\rho}]-\dfrac{1}{\hbar^{2}}[\hat{H}_{1}(t),\int_0^t\text{d}s C(t-s)[\hat{H}_{1}(s),\hat{\rho}(s)]]
\end{align}
where $C(t - s)$ is the correlation function. 
According to Wiener-Khinchin's theory, spectral power density and correlation function are related and the spectral density is equal to the Fourier transform of the correlation function. Among the various types of noises, Gaussian white noise is the most common one. This noise is completely uncorrelated in time and has a flat power spectrum. The dynamics of Markovian can be examined with white noise fluctuations \cite{Aur}. By considering the correlation function $C(t) =\alpha\delta(t)$ for this noise and the corresponding constant power spectrum $ S(\omega) =\dfrac{\alpha}{2\pi} $, master equation 1 simplifies to     
\begin{align}
\label{whitenoise} 
\frac{d}{dt}\hat{\rho}=-\dfrac{i}{\hbar}[\hat{H}_{0},\hat{\rho}]-\dfrac{\alpha}{2\hbar^{2}}[\hat{H}_{1}(t),[\hat{H}_{1}(t),\hat{\rho}(t)]]
\end{align}
where $\alpha$ specifies the noise strength. 
 This equation is obtained by averaging over stochastic classical noises. Using this exact master equation avoids many repetitions to obtain the average dynamics. In Section IV, we will explain the details of using this master equation in our model.
%%%%%%%%%%%%%%%%%%%%%%%%%

\section{Decoherence and Spin–Boson Model }
Every realistic system inevitably interacts with its environment \cite{ala,Hamidr,Carles}. For nanoscale quantum systems and quantum biological systems, this interaction is not insignificant; consequently, they should be considered to be open systems \cite{los,HNaeij}. So, for the dynamic modeling of quantum biological systems, one should go to the theory of open quantum systems. By using decoherence theory and open quantum systems, a realistic picture of a distinct model can be obtained. Due to the manifold interactions between ions and multiple degrees of freedom of the channel, the dynamics of ion transport in the ion channel are very complex. In ion channels, rapid decoherence is apparently inevitable due to the strong coupling of the quantum state of the passing ion with the molecular vibrational modes of the environment of the protein. However, the time of the passage of the ion through the membrane is also very short (about 10-20 ns) \cite{Vaz,Ber}. The ratio of the time scale of the decoherence and the traversal time plays the main role in the understanding of maintaining coherence \cite{Bh}. To maintain the quantum superposition of the transiting ion state, the decoherence time must be longer than this transit time. Based on this, it is suggested that the ion selectivity filter probably has quantum coherence. Note that for more rigorous conclusions, one should wait for more detailed simulations and experimental research. In the interaction between the system and its environment, since the degrees of freedom of the environment are infinite, the system cannot be characterized by specific states or a wave function. T herefore, a density matrix should be attributed to the system and its evolution should be checked over time. The Hamiltonian operator corresponds with energy observations. Therefore, each element of the main diagonal of the density matrix gives the probability of finding the system at a certain energy level \cite{los}. Due to the dissipative terms of the master equation, the time evolution of open systems is non-unitary in contrast to closed systems. The Spin-Boson model has been extensively used in the literature to study dissipation and quantum decoherence. Recently, this model has been used in biological systems to investigate quantum coherence \cite{Gil,Pac,Tao,Hue,Nal}. In this model, a single two-level system, e.g., a spin 1/2 particle, is coupled with a boson environment of harmonic oscillators. In the Spin-Boson model with tunneling, the general form of the Hamiltonian is defined as follows:  
\begin{equation}
\label{spinboson} 
\hat{H}=\frac{1}{2}\omega_{0}\hat{\sigma}_{z}-\frac{1}{2}\Delta_{0}\hat{\sigma}_{x}+\sum_i(\frac{1}{2m_{i}}\hat{p}^{2}_{i}+\frac{1}{2}m_{i}\omega^{2}_{i}\hat{q}^{2}_i)+\hat{\sigma}_{z}\otimes\sum_i c_{i}\hat{q}_{i}
\end{equation} 
where $\Delta_0$ is the tunneling matrix element and represents the tunneling between two quantum states and $c_i$ is the coupling strength of the system-environment.
An appropriate approach to formulating open quantum systems dynamics is to use quantum-dynamical equations, commonly known as quantum master equations (QME). The general expression of the Born-Markov master equation for the Spin-Boson model is obtained as \cite{los}:

\begin{align}
\label{masterequation}
\frac{d}{dt}\hat{\rho}_s(t)=&-\dfrac{i}{\hbar}[\hat{H}^{'}_{s},\hat{\rho}_{s}(t)]-D[\hat{\sigma}_{z},[\hat{\sigma}_{z},\hat{\rho}_{s}(t)]] \nonumber \\
&+\zeta\hat{\sigma}_{z}\hat{\rho}_{s}(t)\hat{\sigma}_{y}+\zeta^{*}\hat{\sigma}_{y}\hat{\rho}_{s}(t)\hat{\sigma}_{z},
\end{align}

where
\begin{equation}
\label{hpe} 
\hat{H}^{'}_{s}=[-\frac{1}{2}\Delta_{0}-\zeta^{*}]\hat{\sigma}_{x}
\end{equation}
is the renormalized (“Lamb-shifted”) Hamiltonian of the system. The coefficient $\zeta$ is provided by

\begin{equation}
\label{zeta}
\zeta=\int_0^{\infty}\text{d}\tau[\nu(\tau)-i\eta(\tau)]\sin(\Delta_{0}\tau)\equiv f-\text{i}\gamma
\end{equation}
The coefficients D, f and $\gamma$ in the above equations are determined as follows:
\begin{equation}
\label{eqD}
D=\int_0^{\infty}\text{d}\tau\nu(\tau)\cos(\Delta_{0}\tau),
\end{equation}
\begin{equation}
\label{eqf}
\textit{f}=\int_0^{\infty}\text{d}\tau\nu(\tau)\sin(\Delta_{0}\tau),
\end{equation}
\begin{equation}
\label{gammaa}
\gamma=\int_0^{\infty}\text{d}\tau\eta(\tau)\sin(\Delta_{0}\tau),
\end{equation}
where
\begin{equation}
\label{ِnu}
\nu(\tau)=\int_0^{\infty}\text{d}\omega J(\omega)\coth(\dfrac{\omega}{2k_{B}T})\cos(\omega\tau)
\end{equation}
and

\begin{equation}
\label{ِeta}
\eta(\tau)=\int_0^{\infty}\text{d}\omega J(\omega)\sin(\omega\tau)
\end{equation}
are the noise and dissipation kernels, respectively \cite{Sha}. Here $J(\omega)$ is the spectral density of the environment and the coupling between the system and the environment is described by it. In this work, an ohmic environment with a Lorentz–Drude cutoff frequency which has high applications, is used. For such environments, the spectral density is written as\cite{Soltanm,Hamid,manesh,manesh}:  
\begin{equation}
\label{ِspec}
J(\omega)=\dfrac{2M\gamma_{0}}{\pi}\omega\dfrac{\omega_{c}^{2}}{\omega_{c}^{2}+\omega^{2}}
\end{equation}
where $\omega_{c}$ is the cutoff frequency and $\gamma_{0}$ constant specifies the effective coupling strength between the system and its environment.
The first term to the right of equation 2 shows the unitary evolution of the density matrix. The second term, which is related to the interaction of the system with the environment, shows decoherence effects and the third and fourth terms describe the decay of the two-level system. In this work, the evolution of the reduced density matrix of the system has been investigated using the master equation 2, which is described in detail in the next section.
%%%%%%%%%%%%%%%%%%%%%%%%%

\section{Model and Methods}
\begin{figure}
\centering
\includegraphics[scale=0.35]{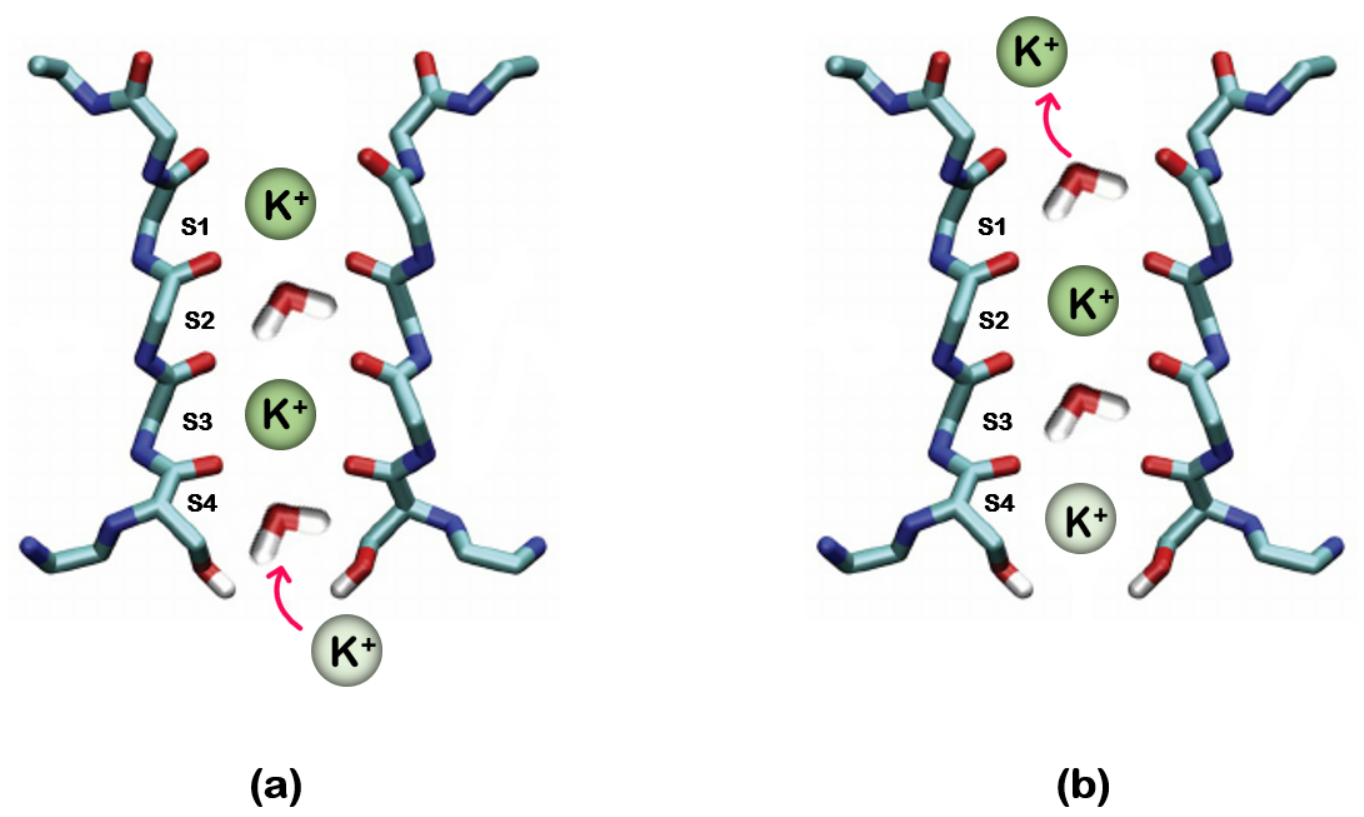}
\caption{Schematic illustration of how a single file is transported through the KcsA ion channel selectivity filter. The two ions move in a concerted manner between two states, (a) 1,3 state and (b) 2,4 state, until a third ion comes in, shifting the ion to the other side of the queue.}
\label{modion}
\end{figure}

Since the width of the selectivity filter is only about 0.3 nm, potassium ions must leave their hydrated shell before entering the selectivity filter and pass-through this filter in a single file \cite{Ben,Corr,Beno}. The crystal structure of the KcsA potassium channel showed that the selectivity filter has four binding sites for $\text{K}^{+}$ (labeled with S1 to S4, from extracellular to intracellular) FIG. \ref{ion}. In order to study the ion permeation mechanism via the selectivity filter, different experimental techniques like radiotracer flux essays and X-ray crystallography, and various computational methods such as molecular dynamics (MD) simulations have been used. The experiments conducted provide support for two mechanisms: "hard-knock" in which water molecules are ignored, and "knock-on" in which water molecules can be either present or absent. To determine the permeation model that is more consistent with the experiment, 2D IR spectra were calculated for all ion configurations using MD simulations. It was found that the knock-on model (with water molecules) is in good agreement with the experiment \cite{Krat}. The knock-on model was first proposed in 1955 by Hodgkin and Keynes \cite{Hod}. The ion transfer in this process is such that the selectivity filter is occupied by two potassium ions within a single file simultaneously with intervening water molecules that oscillate harmoniously between configurations 1 and 3 or 2 and 4 FIG. \ref{modion}. Then the third $\text{K}^{+}$ ion enters from one side and the sequence of ion-water shifts, and finally, the potassium ion is displaced from the opposite side \cite{Ber,Yuf}. It should be noted that the Coulomb interaction is very effective on a very long time scale, so we consider this model on a relatively short time scale. Because $\text{K}^{+}$ ions in the ion channel fluctuate so much, strong Columbian repulsion is probably not present on this time scale. So our model is based on the two described configurations. Therefore, we investigate an ion channel with four sites as shown in FIG. \ref{modion}. In this model, we define a two-state system: (1) $\text{K}^{+}$ ions on the 1 and 3 sites and water molecules on the 2 and 4 sites. (2) $\text{K}^{+}$ ions in the 2 and 4 states and water molecules in the 1 and 3 sites. Jumping from one site to another is associated with an energy barrier. This system can be shown in a two-dimensional Hilbert space with two states of $\vert 0\rangle$ and $\vert 1\rangle$ similar to a spin-1/2 system. In the following stages, we used two Spin-Boson and classical noise models to investigate the decoherence rate in the ion channel.  First, using the Spin-Boson model and master equation \eqref{masterequation}, the evolution of the density matrix was obtained. $\hslash$ is supposed to be 1 in the rest of the paper. 
 In this model, the total Hamiltonian is defined as:
\begin{align}
\label{totalH}
\hat{H}=\hat{H}_s+{\hat{H}_\varepsilon}+\hat{H}_{int}
\end{align}
Here
\begin{align}
\label{totalH}
\hat{H}_s=\dfrac{1}{2}\omega_{0}\hat{\sigma_{z}}-\dfrac{1}{2}\Delta_{0}\hat{\sigma_{x}}
\end{align}
is the self-Hamiltonian of the system. The asymmetry energy $\omega_0$ represents the energy difference between the basis states of the system( the eigenstates of z).

$\hat{H}_\varepsilon$ represents the self-Hamiltonian of the environment which is composed of harmonic oscillators with frequency $\omega$. $\hat{H}_{int}$ describes the linear coupling between the z-coordinate of the system and the position of the coordinates of each environmental harmonic oscillator. According to the above discussion and considering the following equation for

\begin{align}
\label{totalHper}
\hat{H}^{'}_{s}=\dfrac{1}{2}\omega_{0}\hat{\sigma_{z}}-\dfrac{1}{2}\Delta_{0}\hat{\sigma_{x}}-\zeta^{*}\hat{\sigma_{x}}
\end{align}
equation \eqref{masterequation} can be rewritten in the basis of the eigenstates of the $\hat{\sigma_{z}}$ operator as:

\begin{align}
\label{num}
\dfrac{d}{dt}\hat{\rho}_{00}=&\dfrac{i}{2}\Delta_0(\hat{\rho}_{10}-\hat{\rho}_{01})+2\gamma\hat{\rho}_{01} \nonumber \\
\dfrac{d}{dt}\hat{\rho}_{01}=&\dfrac{i}{2}\Delta_0(\hat{\rho}_{11}-\hat{\rho}_{00})+2i\zeta^{*}\hat{\rho}_{11}-2if\hat{\rho}_{00}-(i\omega_{0}-4D)\hat{\rho}_{01} \nonumber \\
\dfrac{d}{dt}\hat{\rho}_{10}=&\dfrac{i}{2}\Delta_0(\hat{\rho}_{00}-\hat{\rho}_{11})+2i\zeta^{*}\hat{\rho}_{00}-2if\hat{\rho}_{11}+(i\omega_{0}-4D)\hat{\rho}_{10} \nonumber \\
\dfrac{d}{dt}\hat{\rho}_{11}=&\dfrac{i}{2}\Delta_0(\hat{\rho}_{01}-\hat{\rho}_{10})+2\gamma\hat{\rho}_{10}
\end{align}
With the condition that $J(\omega)$ is odd, $\gamma $ and $D$ coefficients are simplified to get
\begin{align}
\label{mag}
\gamma=\dfrac{\pi}{2}J(\Delta_{0})
\end{align}
and
\begin{align}
\label{dmag}
D=\dfrac{\pi}{2}J(\Delta_{0})coth(\dfrac{\Delta_{0}}{2k_{B}T})
\end{align}

In the high-temperature limit $\beta\omega_{c}\ll1$( where $\beta=\dfrac{1}{k_{B}T})$, $coth(\dfrac{\beta\omega}{2})\thickapprox2(\beta\omega)^{-1}$ and

\begin{align}
\label{fmag}
f=2M\gamma_{0}k_{B}T\dfrac{\omega_{c}\Delta_{0}}{\Delta_{0}^{2}+\omega_{c}^{2}}
\end{align}

 Now we go to the classical noise formalism and investigate our model using master equation \eqref{whitenoise}. To begin, consider the following total Hamiltonian
\begin{align}
\label{totalHnoise}
\hat{H}(t)=\hat{H}_0+z(t)\hat{H}_{1}
\end{align}
\begin{align}
\label{Hzer}
\hat{H}_{0}=\dfrac{1}{2}\omega_{0}\hat{\sigma_{z}}-\dfrac{1}{2}\Delta_{0}\hat{\sigma_{x}}
\end{align}  
\begin{align}
\label{Hone}
\hat{H}_{1}=\hat{\sigma_{z}}
\end{align}
where $\hat{H}_0 $ is the noise-free Hamiltonian, $\hat{H}_{1}$ describes the “noisy” Hamiltonian, and $z(t)$ is a real function for a given noise realization and it has replaced the interaction of the system-environment. Finally, the master equation \eqref{whitenoise} for Gaussian white noise can be rewritten as follows:

\begin{align}
\label{numm}
\dfrac{d}{dt}\hat{\rho}_{00}=&\dfrac{i}{2}\Delta_0(\hat{\rho}_{10}-\hat{\rho}_{01}) \nonumber \\
\dfrac{d}{dt}\hat{\rho}_{01}=&\dfrac{i}{2}\Delta_0(\hat{\rho}_{11}-\hat{\rho}_{00})-i\omega_{0}\hat{\rho}_{01}-2\gamma\hat{\rho}_{01} \nonumber \\
\dfrac{d}{dt}\hat{\rho}_{10}=&\dfrac{i}{2}\Delta_0(\hat{\rho}_{00}-\hat{\rho}_{11})+i\omega_{0}\hat{\rho}_{10}-2\gamma\hat{\rho}_{10} \nonumber \\
\dfrac{d}{dt}\hat{\rho}_{11}=&\dfrac{i}{2}\Delta_0(\hat{\rho}_{01}-\hat{\rho}_{10})
\end{align}

\begin{center}
\begin{figure*}
\centering
\subfigure[]{\label{Haftreal}
\includegraphics[scale=0.5]{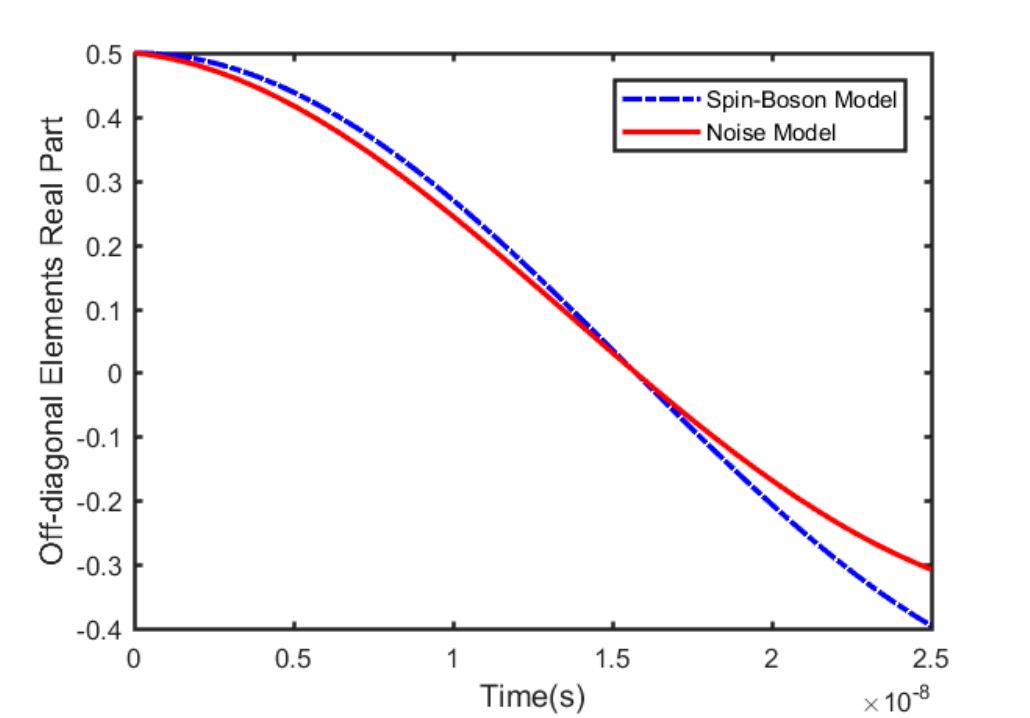}}
\subfigure[]{\label{Haftimag}
\includegraphics[scale=0.5]{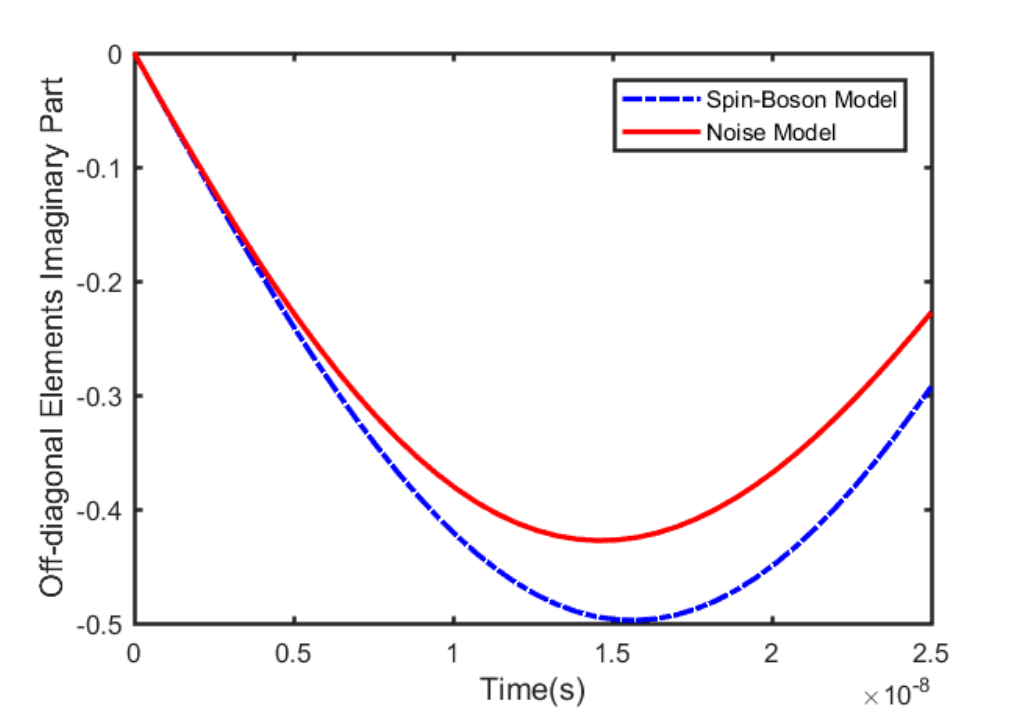}}
\caption {The evolution of off-diagonal elements of density matrix versus time with  $\gamma=0.5\times10^{7}s^{-1}$ and $ \Delta_0=1\times10^{7}s^{-1}$, (a) real parts and (b) imaginary parts.}
\end{figure*}
\end{center}
\begin{center}
\begin{figure*}
\centering
\subfigure[]{\label{Hashtreal}
\includegraphics[scale=0.5]{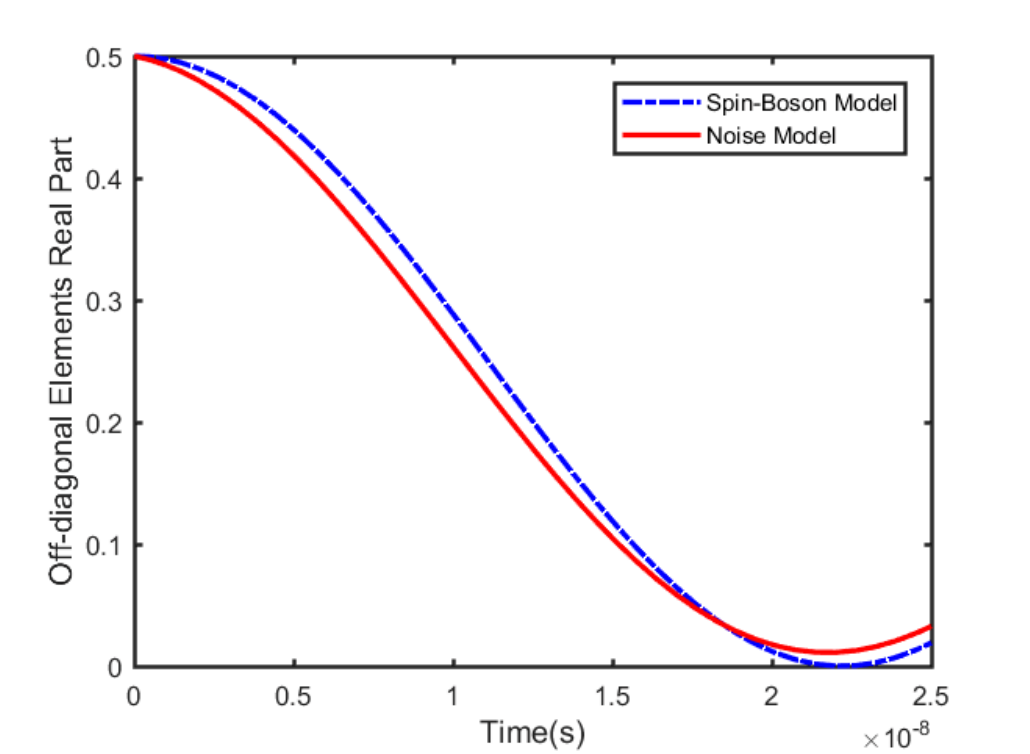}}
\subfigure[]{\label{Hashtimag}
\includegraphics[scale=0.5]{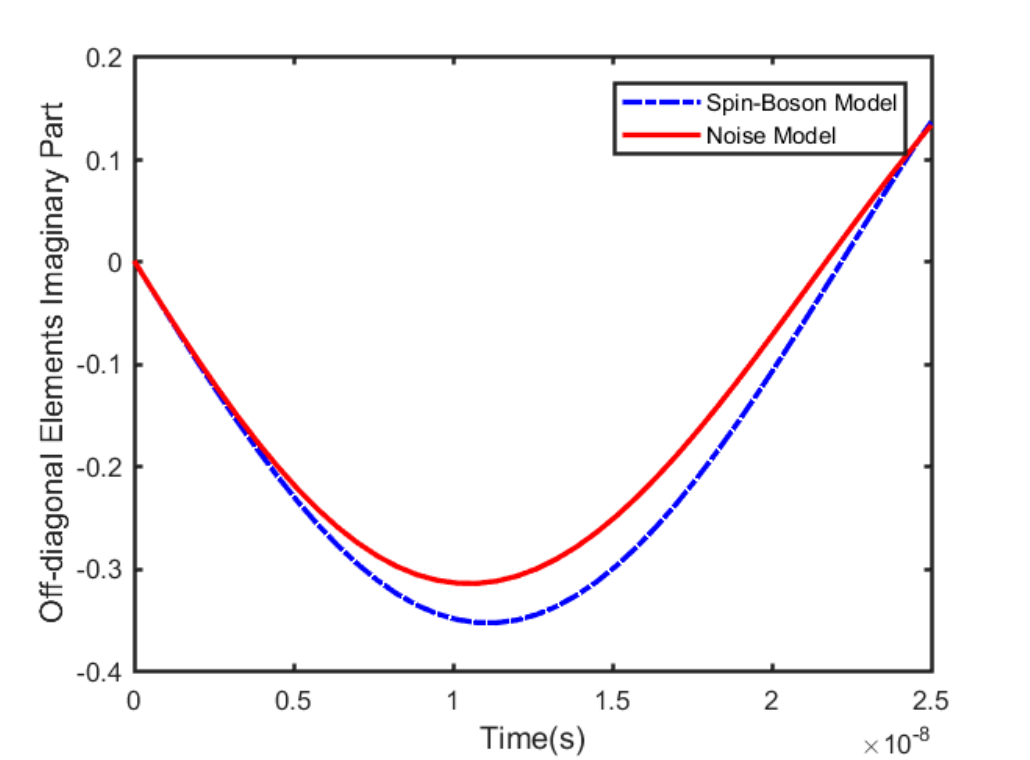}}
\caption {The evolution of off-diagonal elements of density matrix versus time with  $\gamma=0.5\times10^{7}s^{-1}$ and $ \Delta_0=1\times10^{8}s^{-1}$, (a) real parts and (b) imaginary parts.}
\end{figure*}
\end{center}

\begin{figure}
\centering
\includegraphics[scale=0.5]{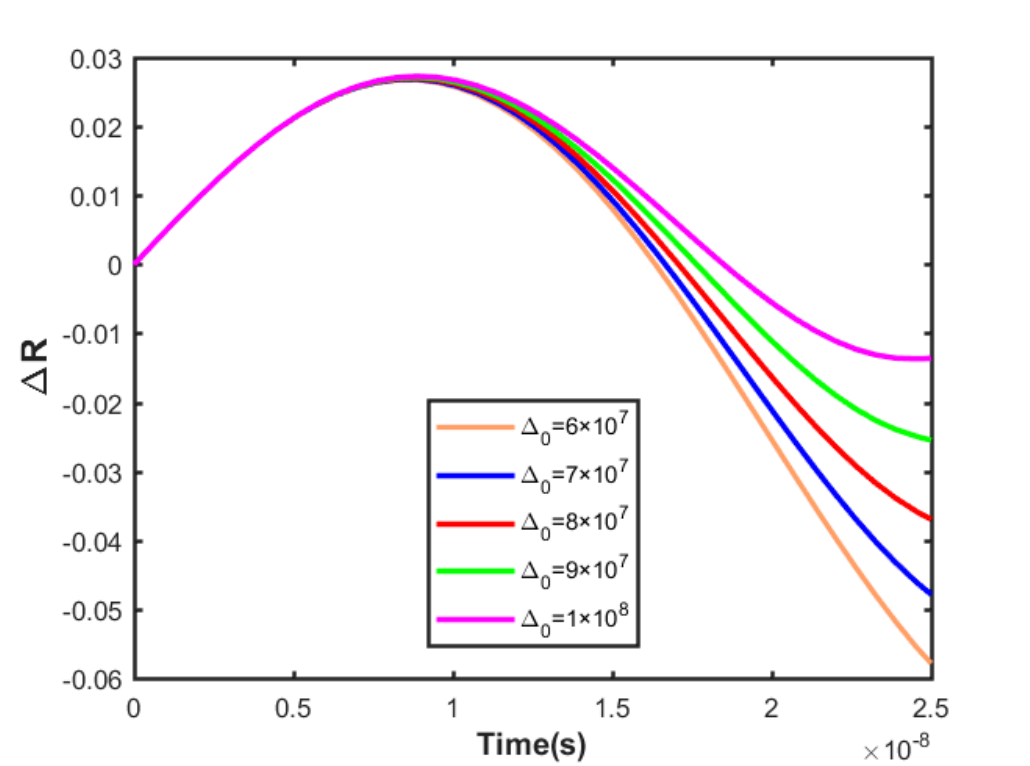}
\caption {The changes of the $\Delta R$ parameter  versus time with  $\gamma=0.5\times10^{7}s^{-1}$ in a variety of hopping rates $ \Delta_0$.}
\label{delta}
\end{figure}

%%%%%%%%%%%%%%%%%%%%%%%%
\section{Results and Discussion}
Recently, it has been suggested that hopping in ion channels may remain coherent during the process \cite{Sei,Vaz,Sal,Sum}. This coherence can play an important role in explaining ion channels' selectivity and ion conduction mechanism. However, based on the quantum mechanical model, this system is expected to rapidly couple with an environment and pass through a decoherence process. Despite the existence of interaction between the channel and the protein environment, quantum coherence is expected to be maintained for shorter time scales than decoherence and play a vital role in the dynamics.
Therefore, in order for the ions to maintain their quantum states during the mechanism of passing through these channels, the decoherence time must be longer than the time required for each ion to pass through the channel. To investigate the problem, first, the system of equation \eqref{masterequation} was solved by considering the initial state of the system as the following superposition:
\begin{align}
\label{is}
\vert\psi_i\rangle=\frac{1}{\sqrt{2}}(|0\rangle+|1\rangle)
\end{align}
Using thermal Broglie wavelength $\lambda_{dB}=1/\sqrt{2mk_BT}$, the order of decoherence in this problem can be obtained at a glance. For the decoherence time one get's
\begin{align}
\label{dtime}
\tau_D=\frac{\Delta X^2}{\gamma \lambda_{dB}^2},
\end{align}
where $\Delta X$ represents the dispersion in position space and
\begin{align}
\label{gamma}
\gamma=\gamma_0\omega\bar{n}\frac{r^2}{1+r^2},
\end{align}
with $r=\Lambda/\omega$ and $\bar{n}=(e^{\omega/k_BT}-1)^{-1}$. $\Lambda$ is the cut-off frequency and $\bar{n}$ demonstrates the mean environmental population based on Temperature ($T$).The value of the diffusing rate strongly depends on the frequency of the particles. The decoherence time for this ion channel based on the selection of the appropriate diffusion rate  ($1\times 10^{6}s^{-1}$ to $1\times 10^{8}s^{-1}$) and its appropriate frequency ($1\times 10^{8}s^{-1}$ to $1\times 10^{12}s^{-1}$)  is about  $1\times 10^{-7}s$ or more. \cite{Sei,Vaz}.
 Considering the selectivity filter as a two-level system and using the Spin-Boson model, the behavior of the system in the interaction with the protein environment was investigated. The coupling between the system and the environment was characterized by the spectral density $J(\omega)$. To check the results of the two models, equations \eqref{num} and \eqref{numm} are solved numerically.  As mentioned, the hopping rate in the potassium ion channel is high and is of $10^{6}$-$10^{8}$ $\text{s}^{-1}$ order. Therefore, the compatibility of these two models is checked at different speeds. Off-diagonal elements show similar behavior, and the probabilities of these elements can be used to check the behavior of the system’s coherence. Here the evolution of the second element is presented for the example. First, these two models were considered at lower speeds, and the results are shown in FIG. 3 for illustration. This figure clearly shows that quantum decoherence results are distanced from the noise model by reducing the speed of passage. This spacing occurs at higher times. Then two models of spin boson and classical noise were investigated at higher speeds and the results are shown in FIG. 4. FIG. 4 clearly shows that with the increase of the ion channel hopping rate, the two models show a very good agreement. By comparing FIGs. 3 and 4, it seems that with the increase in the speed of passing through the channel, the results of the Spin-Boson model approach the classical noise formalism.  In the following, for a more detailed analysis of the results, we defined the $\Delta R$ parameter as follows 

\begin{equation}
\label{dif}
\Delta R=R_{SB}-R_{N},
\end{equation}
where $R_{SB}$ and $R_{N}$, respectively, show the real part of the evolution of the elements off-diagonal of the density matrix for the Spin-Boson model and the noise formalism. $\Delta R$ parameter shows the difference between the results of two Spin-Boson and classical noise models. Therefore, in the following, we use this parameter to compare the results of these models. The results are checked in the range of the speed of the ions passing through the channel. The difference between the results of the two models at different speeds is investigated using equation \eqref{dif}. The results of this equation are shown in FIG. 5. According to this figure, it is clear that the difference between the two models is close to zero at a high jump rate,  and the Spin-Boson model reproduces the classical noise in high hopping rates very well. The results of these two models are distanced as the speed of passage decreases at higher times. It should be noted that over time, the results of the two models diverge, even at high speeds, and at longer times, the difference between the two models is striking. According to FIG 5, it is clear that for higher speeds, for example, $1\times 10^{8}s^{-1}$, the two models have an acceptable match. So, in short, it can be said that the Spin-Boson model is capable of simulating the classical noise model for the ion channel at high speeds. The result of this work is in full agreement with recent discoveries. In a recent study\cite{Sei}, the relationship between the rate of hopping and maintaining coherence in ion channels was investigated and it was found that the system is coherent in ion channels with high throughput rates. The present work also provides further evidence as to why ion channels have a high selectivity rate. It should be noted that for other values different from $\gamma$ and $\omega$ considered here, the two Spin-Boson and classical noise models show good agreement at high speeds. Therefore, the loss of coherence described by the classical noise formalism in the potassium ion channel can be well described by the Spin-Boson model, and these two dynamics are in good agreement. For the mechanisms that ignore the presence of water molecules, the two models have a good match.

 %%%%%%%%%%%%%%%%%%%%%%%%%%%%%%%%%
\section{Conclusion}
The purpose of this article is to investigate the behavior of potassium ion channels by two models: Spin-Boson and classical noise models. Recently, it has been suggested that quantum coherence is part of the ion selection process. But due to the biological temperature and the coupling of the channels with the environment, the system loses its coherence. When the system interacts with its environment, it goes through a decoherence process. In various works, the loss of coherence caused by environmental entanglement has been compared with classical oscillations that cause system disturbances. In other words, classic noise processes are used to simulate the effect of the environment on the system. This is usually done by adding random terms to the Hamiltonian of the system. We compared the quantum decoherence that occurs when a single quantum system is entangled with environmental degrees of freedom with the apparent decoherence obtained by averaging over a set of unitary evolutions produced by a stochastic Hamiltonian in an ion channel system. To investigate the behavior of the ion channel system, we assumed our model consists of the superposition of two states for the system: states 1,3, and  2,4. In the first one, potassium ions are in the 1st and 3rd positions and the 2nd and 4th positions are occupied by water molecules. In the second case, positions 2 and 4 are occupied by potassium ions, and positions 1 and 3 are occupied by water molecules. We studied this process using the two Spin-Boson and classical noise models and solved the equations numerically. At first, this system was considered as a two-level one, which is coupled to the protein environment, which includes a set of coordinated oscillators. The effects of the environment on the system are briefly described by the spectral density. The behavior of this system was investigated by the Spin-Boson model with tunneling. The master equation \eqref{masterequation} was solved and investigated in different regimes. Then, the classical noise model was used to investigate the behavior of the ion channel system. We have shown in what conditions the decoherence caused by the interaction of the ion channel system with its environment can simulate classical noise effects. The results showed that at high hopping rates, the loss of coherence caused by a white Gaussian noise can be well described by the Spin-Boson model considering an ohmic environment with high temperature. By reducing the speed of passage, we can clearly see the distance of quantum decoherence results from the classical noise model. For future work, we can look for a convincing theory and mechanism for the ion channel and investigate the role of quantum coherence in the selectivity process. This mechanism should be able to explain well how an ion channel allows only one specific ion to pass through and rejects all ions of smaller diameters.
%%%%%%%%%%%%%%%%%%%%%%%%%%%%%%%%%
\section*{Additional information}
Correspondence should be addressed to Afshin Shafiee [email: shafiee@sharif.edu].

%%%%%%%%%%%%%%%%%%%%%%%%%%%%%%%%%

\end{document}